\documentclass[12pt]{article}
\title{\bf A Possible Universal Treatment of\\
the Field Strength Correlator\\
in the Abelian-Projected SU(2)-Theory}
\author{Dmitri Antonov \thanks{Permanent address:
Institute of Theoretical and Experimental Physics, 
B. Cheremushkinskaya 25, RU-117 218 Moscow, Russia.}{\,}
\thanks{Tel.: + 39 050 844 536; Fax: + 39 050 844 538; 
E-mail address: {\tt antonov@df.unipi.it}} 
\\
{\it INFN-Sezione di Pisa, Universit\'a degli studi di Pisa,}\\
{\it Dipartimento di Fisica, Via Buonarroti, 2 - Ed. B - 
I-56127 Pisa, Italy}}
\date{}
\begin{document}
\maketitle
\vspace{1mm}
\centerline{\bf {Abstract}}
\vspace{3mm}
\noindent
An integral relation between two functions
parametrizing the bilocal field strength 
correlator within the Stochastic Vacuum Model is obtained 
in the effective Abelian-projected $SU(2)$-theory. This relation is  
independent of the concrete properties of the ensemble of vortex loops, 
which are present in the theory under study. 
By virtue of the lattice result stating that the infrared 
asymptotic behaviours of these functions should have the 
same functional form, the obtained 
relation enables one to find these  
behaviours, as well as the infrared asymptotics 
of the bilocal correlator of 
densities of the vortex loops.  
Those turn out to be exponentials, decreasing 
at the inverse mass of the dual vector boson, times certain polynomials 
in the inverse integer powers of the distance. This result agrees 
with the general predictions and the 
existing lattice data better than the results of previous 
calculations, where these powers were found to be half-integer ones.

\vspace{3mm}

\vspace{3mm}

\newpage

Stochastic Vacuum Model (SVM)~\cite{2} (see~\cite{3} for reviews) 
is nowadays commonly argued to be 
one of the most successful nonperturbative approaches to QCD. 
However, an important problem still requiring its clarification 
is the derivation of the bilocal field strength correlator, which is the 
main quantity in SVM, from the QCD Lagrangian. To simplify this problem 
it is natural to consider such a correlator in Abelian-type models, 
which allow for the analytical description of confinement. In this way, 
in the recent paper~\cite{1}, the bilocal correlator has been 
calculated in the effective Abelian-projected~\cite{thooft} $SU(2)$- and 
$SU(3)$-theories. Those are just the 4D dual Ginzburg-Landau--type 
theories, in which confinement is realized 
as the dual Meissner effect 
according to the 't Hooft-Mandelstam scenario~\cite{scen}.
The main difference of the calculation 
performed in Ref.~\cite{1} {\it w.r.t.} the pure classical calculations  
performed before~\cite{class} was
the account for quantum 
vortex loops present in the above-mentioned  
theories. Being virtual (and therefore small-sized)
objects, these loops have been treated within the 
natural, from the point of view of the standard 
superconductivity~\cite{popov},  
model in which they form a dilute
gas~\cite{gas}. 

The aim of the present Letter is to improve on the 
calculation performed in Ref.~\cite{1} 
and to find the bilocal field strength correlator without any 
assumptions imposed on the ensemble of vortex loops. Such a 
model-independent study may become possible by virtue of the 
lattice data on the SVM correlator~\cite{4, quenched, 5}
(see~\cite{6,7} for recent reviews). The present approach is based on the 
above-mentioned commonly accepted belief that confinement can really
be viewed as the dual superconductivity. Clearly, it is this statement 
which enables one to employ the lattice results on the SVM 
correlator in QCD for the evaluation of an analogous quantity in the 
dual Ginzburg-Landau--type theories. In this respect, it is worth noting 
that recently a new support was 
given to the 't Hooft-Mandelstam scenario of confinement (and consequently 
to the dual Abelian-type models) by the 
evidence of monopole condensation on the lattice~\cite{8} (see
Ref.~\cite{6} for a review).      

In what follows, we shall carry out our analysis for the 
simplest $SU(2)$-case, although the generalization to the $SU(3)$-case
along with the lines of Ref.~\cite{1} is straightforward.
Within the so-called Abelian dominance hypothesis~\cite{abdom}, 
stating that the off-diagonal degrees of freedom are irrelevant to  
confinement, the effective IR theory under study can be shown 
(see {\it e.g.}~\cite{surv}) 
to be nothing else, but the dual Abelian Higgs model with external 
electrically charged particles (quarks). In the London limit,
{\it i.e.} the limit when the mass of the dual Higgs field is much larger than 
the mass of the dual vector boson, the respective action 
reads~\footnote{Throughout the present Letter, all the investigations 
will be performed in the Euclidean space-time.}

\begin{equation}
\label{1}
S=\int d^4x\left[\frac14\left(F_{\mu\nu}+F_{\mu\nu}^{\rm e}\right)^2+
\frac{\eta^2}{2}\left(\partial_\mu\theta-2g_mB_\mu\right)^2\right].
\end{equation}
Here, $\theta$ is the phase of the dual Higgs field describing 
the condensate of monopole Cooper pairs, $\eta$ is the {\it v.e.v.} 
of this field, and $2g_m$ is its magnetic charge with $g_m$ 
being the magnetic coupling constant related to the electric 
one, $g$, as $g_mg=4\pi$. Next, in Eq.~(\ref{1}), $B_\mu$ stands for the 
gauge field dual to the diagonal gluonic field $A_\mu^3$, and 
$F_{\mu\nu}^{\rm e}$ is a field strength tensor of an external quark  
obeying the equation $\partial_\mu\tilde F_{\mu\nu}^{\rm e}(x)=
g\oint\limits_{C}^{}dx_\nu(\tau)\delta(x-x(\tau))$, where 
$\tilde{\cal O}_{\mu\nu}\equiv\frac12\varepsilon_{\mu\nu\lambda\rho}
{\cal O}_{\lambda\rho}$. Note that the field $\theta$ contains 
the multivalued part which describes dual Abrikosov-Nielsen-Olesen 
strings~\cite{ano}. The latter ones can either be open (those end up 
at the contour $C$ and provide the confinement of a quark moving 
along this contour) or closed. Such closed strings with minimal opposite
winding numbers couple to each other and form virtual bound states, 
called vortex loops~\cite{popov, gas}.

Within the SVM, the (irreducible) 
bilocal field strength correlator can be parametrized 
by the two coefficient functions, ${\cal D}\left(x^2\right)$ and 
${\cal D}_1\left(x^2\right)$, 
as follows:

$$\left<\left<f_{\mu\nu}(x)f_{\lambda\rho}(0)
\right>\right>_{A_\mu^3, j_\mu^{\rm m}}=
\Biggl(\delta_{\mu\lambda}\delta_{\nu\rho}-\delta_{\mu\rho}
\delta_{\nu\lambda}\Biggr){\cal D}\left(x^2\right)+$$

\begin{equation}
\label{cumulant}
+\frac12\Biggl[\partial_\mu
\Biggl(x_\lambda\delta_{\nu\rho}-x_\rho\delta_{\nu\lambda}\Biggr)
+\partial_\nu\Biggl(x_\rho\delta_{\mu\lambda}-x_\lambda\delta_{\mu\rho}
\Biggr)\Biggr]{\cal D}_1\left(x^2\right).
\end{equation}
Here, $\left<\left<{\cal O}{\cal O}'\right>\right>\equiv
\left<{\cal O}{\cal O}'\right>-\left<{\cal O}\right>\left<{\cal O}'\right>$,
and $f_{\mu\nu}=\partial_\mu A_\nu^3-\partial_\nu A_\mu^3$. Next,   
the average $\left<\ldots\right>_{A_\mu^3}$ 
in Eq.~(\ref{cumulant}) is just the standard Gaussian 
average over free diagonal gluons, whereas $\left<\ldots
\right>_{j_\mu^{\rm m}}$ is a certain 
average over trajectories of monopole Cooper pairs~\cite{bard}, 
which provides 
pair condensation. Note that it is the coupling of the dual 
field $B_\mu$ to the currents of Cooper pairs 
$j_\mu^{\rm m}$'s, which yields 
nonperturbative contents of the functions ${\cal D}$ and ${\cal D}_1$
in the model under study.

In Ref.~\cite{1}, the following system of equations for the functions 
${\cal D}$ and ${\cal D}_1$ has been obtained:

\begin{equation}
\label{2}
{\cal D}\left(x^2\right)=m^2D_m(x)+\left(4\pi g_m\eta^2\right)^2
\int d^4y\int d^4z
D_m(x-y)D_m(z)\partial^2g(y-z),
\end{equation}

\begin{equation}
\label{3}
G\left(x^2\right)=4D_m(x)+\left(8\pi g_m\eta^2\right)^2
\int d^4y\int d^4z
D_m(x-y)D_m(z)g(y-z).
\end{equation}
Here, $D_m(x)\equiv\frac{m}{4\pi^2|x|}K_1(m|x|)$ is the 
propagator of the dual vector boson of the mass $m$, $m=2g_m\eta$,
$K_\nu$'s henceforth stand for the modified Bessel functions, and 
$G\left(x^2\right)\equiv\int\limits_{x^2}^{+\infty}dt{\cal D}_1(t)$.
Finally, in Eqs.~(\ref{2}) and~(\ref{3}), $g(x)$ denotes a   
scalar function parametrizing the bilocal correlator of densities of 
the vortex loops, $\left<\Sigma_{\mu\nu}(x)\Sigma_{\lambda\rho}(y)
\right>$ (with the average taken over the ensemble of these 
loops), as follows: $\left<\Sigma_{\mu\nu}(x)\Sigma_{\lambda\rho}(y)
\right>=\varepsilon_{\mu\nu\alpha\beta}\varepsilon_{\lambda\rho\gamma\beta}
\partial_\alpha^x\partial_\gamma^yg(x-y)$. 
Although the tensor 
structure of this expression is unambiguously fixed by the condition  
of closeness 
of the vortex loops, $\partial_\mu\Sigma_{\mu\nu}=0$, the form of the 
function $g(x)$ depends on the properties of the ensemble of 
these objects.~\footnote{Note once more that the classical, 
{\it i.e.} $g(x)$-independent, parts of 
Eqs.~(\ref{2}) and~(\ref{3}) have been found in Ref.~\cite{class}.} 
(In particular, in Ref.~\cite{1}, this function has been 
found in the framework of a dilute gas model for the ensemble of 
vortex loops.) 
As we shall see below, the 
IR asymptotics of the function $g(x)$ 
can be found without any model assumptions, 
but rather on the basis of the statement that the respective 
asymptotics of the ${\cal D}$- and 
${\cal D}_1$-functions have the same fall off, as 
it is suggested by the lattice data~\cite{4,quenched,5,6,7}.

To proceed with the study of the system~(\ref{2})-(\ref{3}), notice  
that the integrals over $z$ standing in these equations can be carried out 
[in Eq.~(\ref{2}), by doing firstly the double partial integration] by virtue 
of the formula $\int d^4uD_m(x-u)D_m(y-u)=\frac{1}{8\pi^2}
K_0(m|x-y|)$.~\footnote{The details of a derivation of a more general  
formula, where the masses on the L.H.S. are different, can be found 
in the Appendix to Ref.~\cite{1}.} This yields

\begin{equation}
\label{4}
{\cal D}\left(x^2\right)=m^2D_m(x)+\left(4\pi g_m\eta^2\right)^2
\int d^4y\left[\frac{m^2}{8\pi^2}K_0(m|x-y|)-D_m(x-y)\right]g(y),
\end{equation}

\begin{equation}
\label{5}
G\left(x^2\right)=4D_m(x)+8g_m^2\eta^4\int d^4yK_0(m|x-y|)g(y).
\end{equation}
Differentiating Eq.~(\ref{5}) {\it w.r.t.} $x^2$, one can 
completely eliminate the 
$g(x)$-dependence from the resulting system of equations without 
solving them {\it w.r.t.} this function. This yields the following 
relation between ${\cal D}$ and ${\cal D}_1$, which is thus 
independent of the 
properties of the ensemble of vortex loops:

\begin{equation}
\label{7}
{\cal D}\left(x^2\right)+{\cal D}_1\left(x^2\right)+\frac{m^2}{4}
\int\limits_{x^2}^{+\infty}dt{\cal D}_1(t)=
\frac{2}{x^2}D_m(x)+
\left(\frac{m}{2\pi|x|}\right)^2(K_0(m|x|)+K_2(m|x|)).
\end{equation}

By imposing some relation on the IR asymptotics 
${\cal D}^{\rm as}$ and ${\cal D}_1^{\rm as}$ 
of the functions 
${\cal D}$ and ${\cal D}_1$, one can employ Eq.~(\ref{7}) in order to find 
these asymptotics explicitly. 
The simplest relation of this kind can be imposed  
by disregarding the perturbative-type contributions to 
${\cal D}^{\rm as}$ and ${\cal D}_1^{\rm as}$,
which yields ${\cal D}_1^{\rm as}=\alpha {\cal D}^{\rm as}$ 
with $\alpha\simeq 0.3$~\cite{quenched}. Implementing this 
relation into Eq.~(\ref{7}) and differentiating the IR asymptotics of that 
equation {\it w.r.t.} $x^2$ we get

\begin{equation}
\label{F}
\frac{d{\cal D}^{\rm as}\left(\xi^2\right)}{d\xi^2}-
a{\cal D}^{\rm as}\left(\xi^2\right)=\frac{1}{\alpha+1}F(\xi),
\end{equation}
where $F(\xi)\equiv
-\frac{m^4}{4\sqrt{2}
\pi^{3/2}\xi^{7/2}}\left(1+\frac{3}{\xi}+\frac{3}{\xi^2}\right)
{\rm e}^{-\xi}$ with 
$\xi\equiv m|x|\gg1$ and $a\equiv\frac{\alpha}{4(\alpha+1)}$. 
The solution to this equation, vanishing at infinity, obviously reads

\begin{equation}
\label{int}
{\cal D}^{\rm as}\left(\xi^2\right)=
\frac{m^4}{4\sqrt{2}\pi^{3/2}
(\alpha+1)}{\rm e}^{a\xi^2}\int\limits_{\xi^2}^{+\infty}
\frac{dt}{t^{7/4}}\left(1+\frac{3}{\sqrt{t}}+\frac{3}{t}\right)
{\rm e}^{-at-\sqrt{t}}.
\end{equation}
The details of calculation of the last integral up to the terms 
of the order of $1/\xi$ are outlined in the Appendix, and the result 
has the form

\begin{equation}
\label{fin}
{\cal D}^{\rm as}\left(\xi^2\right)=
\frac{2\sqrt{2}a^{3/4}m^4}{\pi^{3/2}(\alpha+1)}\left[C+
\frac{1}{8a\xi}\left(3C_1+5C_2+7C_3\right)+O
\left(\frac{1}{\xi^2}\right)\right]{\rm e}^{-\xi}.
\end{equation}
The coefficients here read: $C_1=\frac{1}{21\Gamma(3/4)}$, 
$C_2=\frac{4\sqrt{a}}{15\Gamma(1/4)}$, 
$C_3=\frac{4a}{77\Gamma(3/4)}$, $C=C_1+C_2+C_3$ with ``$\Gamma$'' 
denoting the gamma function. Thus we see that the lattice-inspired 
suggestion that the IR asymptotic behaviours  
of both functions ${\cal D}$ and ${\cal D}_1$ are proportional to each other 
yields for those 
the exponential fall off at the inverse mass of the dual vector boson 
times some polynomial in the inverse integer powers of the distance. 
Such a preexponential behaviour differs from that of the classical 
calculation~\cite{class} and the one which was obtained in the 
dilute gas model of the ensemble of vortex loops~\cite{1}.
Indeed, in that cases the preexponentials were half-integer 
inverse powers of the distance, which is less favourable from the 
point of view of the lattice data~\cite{4,quenched,5,6,7}.

Let us now investigate the stability of the obtained solution~(\ref{fin})
{\it w.r.t.} some possible power-like corrections. Namely, let us 
insert into Eq.~(\ref{F}) the following modified {\it Ansatz}:

\begin{equation}
\label{modif}
{\cal D}_1^{\rm as}\left(\xi^2\right)=
\alpha\left(\xi^2\right)^{-\lambda}{\cal D}^{\rm as}\left(\xi^2\right)
\end{equation}
with $\lambda\to 0$. In the limit $\xi\gg1$ under study this leads
to the equation 

$$
\left[\frac{1}{\alpha}\left(\xi^2\right)^\lambda+1\right]
\frac{d{\cal D}_1^{\rm as}}{d\xi^2}-\frac14{\cal D}_1^{\rm as}
=F(\xi),$$
or equivalently 

\begin{equation}
\label{eps}
\frac{d{\cal D}_1^{\rm as}}{d\xi^2}-a(1+\varepsilon){\cal D}_1^{\rm as}
=4a(1+\varepsilon)F(\xi),
\end{equation}
where $\varepsilon\equiv\frac{1-\left(\xi^2\right)^\lambda}{\alpha+1}$.
In what follows we assume that $\varepsilon\ll 1$, which is obviously 
true for sufficiently small $\lambda$, and 
seek for the solution to Eq.~(\ref{eps}) in the form 
${\cal D}_1^{\rm as}={\cal D}_1^{{\rm as}{\,}(0)}+\varepsilon
{\cal D}_1^{{\rm as}{\,}(1)}$. Thus, to check the stability of the 
obtained solution~(\ref{fin}), one should find ${\cal D}_1^{{\rm as}{\,}(1)}$
and prove that it is not dramatically large {\it w.r.t.} 
${\cal D}_1^{{\rm as}{\,}(0)}$. The leading term of the latter one 
reads $B{\rm e}^{-\xi}$, where
$B\equiv\frac{8\sqrt{2}a^{7/4}Cm^4}{\pi^{3/2}}$, and for the desired
function ${\cal D}_1^{{\rm as}{\,}(1)}$ we get the following equation:

$$
\frac{d{\cal D}_1^{{\rm as}{\,}(1)}}{d\xi^2}-a{\cal D}_1^{\rm as{\,}(1)}=
4a\left[
F+\frac14{\cal D}_1^{{\rm as}{\,}(0)}\right]\simeq aB{\rm e}^{-\xi}.$$
This equation is straightforward to be integrated, and the resulting 
integral can be evaluated in the large-$\xi$ limit by virtue of the 
known asymptotics for the probability integral~\cite{gr}: 
$\frac{2}{\sqrt{\pi}}\int\limits_{0}^{x}dt{\rm e}^{-t^2}\to 
1-\frac{1}{\sqrt{\pi}}\frac{{\rm e}^{-x^2}}{x}$ at $x\gg1$. In this way, 
we obtain ${\cal D}_1^{{\rm as}{\,}(1)}= 
-{\cal D}_1^{{\rm as}{\,}(0)}$,
and thus finally ${\cal D}_1^{\rm as}=B(1-\varepsilon)
{\rm e}^{-\xi}$. We conclude that while $\varepsilon\ll 1$, {\it i.e.} 
for small enough $\lambda$,
namely $\lambda\ll\left(\ln\xi^2\right)^{-1}$, contributions to the 
obtained solution~(\ref{fin}) stemming 
from the power-like correction~(\ref{modif}) are small, and thus 
this solution is stable.

It is also worth making the following comment. In previous calculations, 
the functions ${\cal D}\left(x^2\right)$ and $G\left(x^2\right)$ 
were obviously proportional to some propagators. For example,  
in the classical approximation those were just $D_m$
[see the first terms on the R.H.S.'s of Eqs.~(\ref{2})-(\ref{3})].
However, this fact is not valid any more in the present approach.
While this statement 
is obvious for the ${\cal D}$-function whose IR asymptotics is given by 
Eq.~(\ref{fin}), some comment is in order for the function 
$G\left(x^2\right)$. Namely, we should check that the leading term 
in the IR asymptotics 
of this function differs from the respective asymptotics of 
any propagator. In this way, we have

\begin{equation}
\label{asG}
G\left(x^2\right)\stackrel{|x|\gg m^{-1}}{\longrightarrow}
\frac{8\sqrt{2}Ca^{7/4}m^4}{\pi^{3/2}}\int\limits_{x^2}^{+\infty}
dt{\rm e}^{-m\sqrt{t}}\simeq\frac{16\sqrt{2}Ca^{7/4}m^2}{\pi^{3/2}}
\xi {\rm e}^{-\xi},
\end{equation}
which explicitly proves our statement\footnote{Moreover, assuming 
that Eqs.~(\ref{fin}) and~(\ref{asG}) remain 
valid also for $|x|\le m^{-1}$, one can explicitly see 
that the Fourier transforms of these functions possess 
only cuts rather than poles. Indeed, in that case we have 
for the leading terms of the ${\cal D}$- and $G$-functions:

$$
{\cal D}\left(p^2\right)\simeq\frac{2\sqrt{2}Ca^{3/4}m^4}{\pi^{3/2}
(\alpha+1)}\int d^4x{\rm e}^{-m|x|+ipx}=
\frac{24\sqrt{2\pi}Cm^5}{\alpha+1}\frac{2m^2-3p^2}{(p^2+m^2)^{7/2}},$$

$$G\left(p^2\right)\simeq 
\frac{16\sqrt{2}Ca^{7/4}m^3}{\pi^{3/2}}\int d^4x|x|{\rm e}^{-m|x|+ipx}=
192\sqrt{2\pi}Ca^{7/4}m^3\frac{3p^4-24m^2p^2+8m^4}{(p^2+m^2)^{9/2}}.$$}.
The fact that in the present approach the functions ${\cal D}$ and $G$
are not proportional to any (massive) propagator makes it closer to
QCD than the previous ones. Indeed, would such a proportionality 
takes place in QCD, it might 
be suspicious of the appearance of an asymptotic state carrying 
colour in some processes involving the bilocal field strength correlator. 
   
Another comment which is in order concerns the comparison of the 
preexponent in Eq.~(\ref{fin}) with the respective value known from the 
lattice measurements. In this way, we immediately face the following problem.
Namely, in our calculation we employed the assumption that 
${\cal D}^{\rm as}\sim{\cal D}_1^{\rm as}$ at $\xi\gg1$.
This relation has been checked on the lattice for the quenched 
$SU(3)$ QCD in Ref.~\cite{quenched} within the interval of distances
between 0.4 fm and 1 fm, which corresponds to the values of 
$\xi$ in the range between 1.8 and 4.6. However, it is straightforward to 
see that the $\frac{1}{\xi}$-term on the R.H.S. of Eq.~(\ref{fin})
becomes smaller than $C$ only for $\xi\ge 5.3$. This means that
the assumption on the proportionality of 
${\cal D}^{\rm as}$ and ${\cal D}_1^{\rm as}$, which led to Eq.~(\ref{fin}),
should be extended to the distances exceeding those which were used 
up to now in the lattice measurements. Thus, since these measurements yield
the preexponential factor at the distances where the next-to-constant 
terms on the R.H.S. of Eq.~(\ref{fin}) are important, the only way
to get this factor within our approach is to integrate numerically 
the exact (rather than the asymptotic) equation for the function ${\cal D}$
down to that distances.
Such an equation is given by the formula~(\ref{F}) with the asymptotic 
function $F$ replaced by the exact one:

$$
-\frac{m^4}{4\pi^2\xi^3}\left[\frac{3K_1(\xi)}{\xi^2}+\frac{3}{2\xi}
(K_0(\xi)+K_2(\xi))+\frac14(3K_1(\xi)+
K_3(\xi))\right].
$$
By virtue of the MATHEMATICA program, one gets for the quantity
${\cal D}\left(\xi^2\right){\rm e}^\xi/m^4$  
at $|x|=0.4{\,}{\rm fm}$ 
the value 0.03, which differs in one order from the lattice 
result~\cite{quenched} equal to 0.32(3). Clarification of 
the origin of this numerical discrepancy requires more precise 
investigations, which will be performed in future publications.

Finally, by virtue of the obtained results it is also 
possible to derive the 
IR asymptotics of the 
bilocal correlator of densities of the 
vortex loops, {\it i.e.} the function $g(x)$.
To this end, let us apply the operator 
$\left(\partial^2-m^2\right)$ to both sides of Eq.~(\ref{5})
differentiated {\it w.r.t.} $x^2$, which yields

$$g(x)\stackrel{|x|\gg m^{-1}}{\longrightarrow}\frac{1}{(2\pi\eta)^2}
\left(\frac{\partial^2}{\partial\xi_\mu^2}-1\right)\left[
\frac{2}{x^2}D_m(x)+\left(\frac{m}{2\pi|x|}\right)^2(K_0(\xi)+K_2(\xi))-
\right.$$

$$\left.-\frac{8\sqrt{2}a^{7/4}m^4}{\pi^{3/2}}
\left(C+\frac{1}{8a\xi}
(3C_1+5C_2+7C_3)+
O\left(\frac{1}{\xi^2}\right)\right){\rm e}^{-\xi}\right],$$
where $\xi_\mu\equiv mx_\mu$. Taking into account that 
$\frac{\partial^2}{\partial\xi_\mu^2}f(\xi)=\frac{3}{\xi}f'+f''$, 
we finally obtain 

$$g(\xi)\stackrel{\xi\gg 1}{\longrightarrow}
\frac{8\sqrt{2}a^{7/4}(g_mm)^2}{\pi^{7/2}}\left[
\frac{3C}{\xi}+O\left(\frac{1}{\xi^2}\right)\right]
{\rm e}^{-\xi}.$$
This IR behaviour of the function $g(x)$ differs from the one obtained 
in Ref.~\cite{1} within the dilute gas model for the ensemble of 
vortex loops. In particular, the obtained expression does not contain 
the asymptotics of the massless propagator, 
which in that case was the origin of a 
novel nonperturbative $1/|x|^4$-term in the ${\cal D}_1$-function. 
In fact, this term in the function ${\cal D}_1$ is now absent, 
which could be anticipated beforehand owing to the equation 
${\cal D}_1^{\rm as}=\alpha {\cal D}^{\rm as}$. 
Clearly, due to this equation, if 
such a term was present in the function ${\cal D}_1$, it would 
unavoidably appear in the function ${\cal D}$ as well. The latter 
fact would however contradict the general principles of SVM~\cite{2,3}, 
which state that the nonperturbative part of the function 
${\cal D}$ cannot contain any $1/|x|^4$-term.

In conclusion of the present Letter, 
on the basis of the 't Hooft-Mandelstam scenario
of confinement [which suggests that the dual Abelian Higgs model 
is relevant to the description of confinement in the $SU(2)$-QCD] 
and the lattice result on the 
bilocal field strength correlator in QCD (which states that the 
IR asymptotic behaviours of the two 
coefficient functions parametrizing this correlator in SVM have the same 
exponential form), we have 
derived the IR asymptotics of 
this quantity in the London limit of the 
dual Abelian Higgs model. In this way, 
the demand of correspondence with 
the above-mentioned lattice result enabled not to employ any particular 
model of the ensemble of vortex loops, present in the theory under 
study. It rather occurred that the proposed approach yielded as 
a by-product the bilocal correlator of densities of the 
vortex loops itself. As far as the 
field strength correlator is concerned, it turned out to 
decrease exponentially at the inverse mass of the dual vector 
boson with the preexponential given by a certain polynomial in the 
inverse powers of the distance. These powers were found to be 
integer ones, which is in the 
better agreement with the existing lattice data than the half-integer 
powers found in the previous calculations. Besides that, the 
obtained leading terms in the IR asymptotic behaviours of the
functions describing the surface and contour exchanges by means 
of the bilocal field strength correlator have been found to be 
different from that of the massive propagator. This result is
more favourable from the point of view of QCD than the opposite one
obtained in the previous approaches.

\section*{Acknowledgments}

The author is very greatful to Prof. A. Di Giacomo for many
helpful suggestions and discussions, and cordial hospitality.
He is also indebted for useful discussions 
to Profs. H.G. Dosch, D. Gromes, and M.G. Schmidt and to 
Drs. N. Brambilla, E. Meggiolaro, C. Schubert, and A. Vairo.
And last but not least, the author acknowledges the staff 
of the Quantum Field Theory Division
of the University of Pisa for kind hospitality and INFN for  
the financial support.

\section*{Appendix. Calculation of the integral~(\ref{int})}

In this Appendix, we shall present some details of calculation 
of the integral standing on the R.H.S. of Eq.~(\ref{int}). 
Let us start with the contribution 
to this integral brought about by the 
addendum dominant in the large-$\xi$ limit under study,

$$\int\limits_{\xi^2}^{+\infty}\frac{dt}{t^{7/4}}{\rm e}^{-at-
\sqrt{t}}=\frac{1}{\Gamma(7/4)}\int\limits_{0}^{+\infty}
d\tau\tau^{3/4}\int\limits_{\xi^2}^{+\infty}dt{\rm e}^{-t(\tau+a)-
\sqrt{t}}=$$

$$=\frac{1}{\Gamma(7/4)}\int\limits_{0}^{+\infty}
d\tau\tau^{3/4}\int\limits_{0}^{+\infty}\frac{d\lambda}{\sqrt{\pi
\lambda}}{\rm e}^{-\lambda}\int\limits_{\xi^2}^{+\infty}dt
{\rm e}^{-t(\tau+A)}=
\frac{1}{\Gamma(7/4)}
\int\limits_{0}^{+\infty}\frac{d\lambda}{\sqrt{\pi
\lambda}}{\rm e}^{-\lambda}\int\limits_{0}^{+\infty}d\tau\tau^{3/4}
\frac{{\rm e}^{-(\tau+A)\xi^2}}{\tau+A},\eqno(A.1)$$
where we have denoted for brevity $a+\frac{1}{4\lambda}$ by $A$.
The two other addendums can be treated analogously:

$$
3\int\limits_{\xi^2}^{+\infty}\frac{dt}{t^{9/4}}{\rm e}^{-at-
\sqrt{t}}=\frac{3}{\Gamma(9/4)}
\int\limits_{0}^{+\infty}\frac{d\lambda}{\sqrt{\pi
\lambda}}{\rm e}^{-\lambda}\int\limits_{0}^{+\infty}d\tau\tau^{5/4}
\frac{{\rm e}^{-(\tau+A)\xi^2}}{\tau+A},\eqno(A.2)$$

$$
3\int\limits_{\xi^2}^{+\infty}\frac{dt}{t^{11/4}}{\rm e}^{-at-
\sqrt{t}}=\frac{3}{\Gamma(11/4)}
\int\limits_{0}^{+\infty}\frac{d\lambda}{\sqrt{\pi
\lambda}}{\rm e}^{-\lambda}\int\limits_{0}^{+\infty}d\tau\tau^{7/4}
\frac{{\rm e}^{-(\tau+A)\xi^2}}{\tau+A}.\eqno(A.3)$$

One can further write
down the following dominant contribution to the integral over $\tau$
in Eq.~(A.1):

$$\frac{{\rm e}^{-A\xi^2}}{A}
\int\limits_{0}^{A}d\tau\tau^{3/4}+
\int\limits_{A}^{+\infty}d\tau
\frac{{\rm e}^{-\tau\xi^2}}{\tau^{1/4}}=
\left[\frac47A^{3/4}+O\left(\frac{1}{\xi^2}\right)\right]
{\rm e}^{-A\xi^2}.\eqno(A.4)
$$
Here, we have used the following asymptotics of the 
incomplete gamma function at large values of its second 
argument~\cite{gr}: 
$\Gamma(c, z)=z^{c-1}{\rm e}^{-z}\left[1+
O\left(\frac{1}{z}
\right)\right]$, $z\gg 1$. 
The integrals over $\tau$ entering Eqs.~(A.2) and (A.3) can be evaluated 
analogously and read

$$\left[\frac49A^{5/4}+O\left(\frac{1}{\xi^2}\right)\right]
{\rm e}^{-A\xi^2}\eqno(A.5)$$ 
and

$$\left[\frac{4}{11}A^{7/4}+O\left(\frac{1}{\xi^2}\right)\right]
{\rm e}^{-A\xi^2},\eqno(A.6)$$
respectively. Inserting now Eqs.~(A.4)-(A.6)
into Eqs.~(A.1)-(A.3) 
and using the formula $\Gamma(z+1)=z\Gamma(z)$
we get from the original Eq.~(\ref{int}):

$${\cal D}\left(\xi^2\right)=\frac{2\sqrt{2}m^4}{\pi^{3/2}
(\alpha+1)}\int\limits_{0}^{+\infty}\frac{d\lambda}{\sqrt{\pi\lambda}}
{\rm e}^{-\lambda}\left[
\frac{1}{21\Gamma(3/4)}
\left(a-\frac{\partial}{\partial\xi^2}\right)^{3/4}
+\frac{4}{15\Gamma(1/4)}\left(a-
\frac{\partial}{\partial\xi^2}\right)^{5/4}+
\right.$$

$$
\left.+\frac{4}{77\Gamma(3/4)}
\left(a-\frac{\partial}{\partial\xi^2}\right)^{7/4}+
O\left(\frac{1}{\xi^2}\right)\right]
{\rm e}^{-\frac{\xi^2}{4\lambda}}=$$

$$=\frac{2\sqrt{2}a^{3/4}m^4}{\pi^{3/2}
(\alpha+1)}\left[C_1\left(1-\frac1a\frac{\partial}{\partial\xi^2}
\right)^{3/4}+C_2\left(1-\frac1a\frac{\partial}{\partial\xi^2}
\right)^{5/4}+C_3\left(1-\frac1a\frac{\partial}{\partial\xi^2}
\right)^{7/4}+O\left(\frac{1}{\xi^2}\right)\right]{\rm e}^{-\xi}.$$
The constants $C_1$, $C_2$, and $C_3$ entering this result, are introduced 
after Eq.~(\ref{fin}). Finally, taking into account that 
$\left(1-\frac1a\frac{\partial}{\partial\xi^2}\right)^q{\rm e}^{-\xi}=
\left[1+\frac{q}{2a\xi}+O\left(\frac{1}{\xi^2}\right)\right]
{\rm e}^{-\xi}$, 
we arrive at Eq.~(\ref{fin}) of the main text.


\newpage


\begin{thebibliography}{99}


\bibitem{2}
H.G. Dosch, Phys. Lett. {\bf B 190} (1987) 177; 
Yu.A. Simonov, Nucl. Phys. {\bf B 307} (1988) 512;
H.G. Dosch and Yu.A. Simonov, Phys. Lett. {\bf B 205} (1988) 339.

\bibitem{3}
H.G. Dosch, Prog. Part. Nucl. Phys. {\bf 33} (1994) 121;
Yu.A. Simonov, Phys. Usp. {\bf 39} (1996) 313; 
H.G. Dosch, V.I. Shevchenko, and Yu.A. Simonov, 
preprint {\tt hep-ph/0007223} (2000).

\bibitem{1}
D. Antonov, JHEP {\bf 07} (2000) 055.

\bibitem{thooft}
G. 't Hooft, Nucl. Phys. {\bf B 190} (1981) 455.

\bibitem{scen}
S. Mandelstam, Phys. Rep. {\bf C 23} (1976) 245; 
G. 't Hooft, in: {\it High Energy Physics, Vol. 2}, Ed. A. Zichichi 
(Editrice Compositori, Bologna, 1976), p. 1225.
 

\bibitem{class}
D.V. Antonov, Mod. Phys. Lett. {\bf A 13} (1998) 659;
M. Baker, N. Brambilla, H.G. Dosch, and A. Vairo, Phys. Rev. {\bf D 58} 
(1998) 034010; U. Ellwanger, Eur. Phys. J. {\bf C 7} (1999) 673;
D. Antonov and D. Ebert, Phys. Lett. {\bf B 444} (1998) 208;
Eur. Phys. J. {\bf C 8} (1999) 343.

\bibitem{popov}
V.N. Popov, {\it Functional Integrals in Quantum Field Theory and 
Statistical Physics} (Reidel, Dordrecht, 1983).

\bibitem{gas}
D. Antonov, Int. J. Mod. Phys. {\bf A 14} (1999) 4347; 
Mod. Phys. Lett. {\bf A 14} (1999) 1829.

\bibitem{4}
M. Campostrini, A. Di Giacomo, and G. Mussardo, Z. Phys. {\bf C 25} 
(1984) 173.

\bibitem{quenched}
A. Di Giacomo and H. Panagopoulos, Phys. Lett. {\bf B 285} (1992) 133.

\bibitem{5}
L. Del Debbio, A. Di Giacomo, and Yu.A. Simonov, 
Phys. Lett. {\bf B 332} (1994) 111; M. D'Elia, A. Di Giacomo, and 
E. Meggiolaro, Phys. Lett. {\bf B 408} (1997) 315; A. Di Giacomo, 
E. Meggiolaro, and H. Panagopoulos, Nucl. Phys. {\bf B 483} (1997) 371.


\bibitem{6}
A. Di Giacomo, preprint {\tt hep-lat/9912016} (1999); 
preprint {\tt hep-lat/0012013} (2000). 

\bibitem{7}
E. Meggiolaro, Phys. Lett. {\bf B 451} (1999) 414.

\bibitem{8}
A. Di Giacomo, B. Lucini, L. Montesi, and G. Paffuti, 
Phys. Rev. {\bf D 61} (2000) 034503; ibid. {\bf D 61} (2000) 034504.


\bibitem{abdom}
Z.F. Ezawa and A. Iwazaki, Phys. Rev. {\bf D 25} (1982) 2681; 
ibid. {\bf D 26} (1982) 631.

\bibitem{surv}
D. Antonov, Surveys High Energ. Phys. {\bf 14} (2000) 265.

\bibitem{ano}
A.A. Abrikosov, Sov. Phys. JETP {\bf 5} (1957) 1174; 
H.B. Nielsen and P. Olesen, Nucl. Phys. {\bf B 61} (1973) 45.

\bibitem{bard}
K. Bardakci and S. Samuel, Phys. Rev. {\bf D 18} (1978) 2849.

\bibitem{gr}
I.S. Gradshteyn and I.M. Ryzhik, {\it Table of Integrals, Series, and 
Products} (Academic Press, Inc., Orlando, 1980).




\end{thebibliography}
\end{document}